\magnification=\magstep1
\tolerance=500
\vskip 2true cm
\rightline{TAUP 2728-2002}
\rightline{30 December, 2002}
\bigskip
\centerline{\bf Relativistic Brownian Motion}
\bigskip
\centerline{O. Oron and L.P. Horwitz\footnote{*}{Also at Department of
Physics, Bar Ilan University,  Ramat Gan 529000, Israel}}
\smallskip
\centerline{School of Physics and Astronomy}
\centerline{Raymond and Beverly Sackler Faculty of Exact Sciences}
\centerline{Tel Aviv University, Ramat Aviv 69978, Israel}
\bigskip
\noindent {\it Abstract:\/}
\par We solve the problem of formulating Brownian motion in a
relativistically covariant framework in $1+1$ and $3+1$ dimensions.
We obtain covariant Fokker-Planck equations with (for the isotropic
case) a differential operator of invariant d'Alembert form. Treating the
spacelike and timelike fluctuations separately, we show that it is
essential to take into account the analytic continuation of 
``unphysical'' fluctuations in order to achieve these results.

\vfill
\break
\bigskip
\noindent{\bf Introduction}
\par Nelson$^1$, in 1966, constructed the Schr\"odinger equation from
an analysis of Brownian motion by identifying the forward and backward
average velocities of a Brownian particle with the real and imaginary
parts of a wave function. He pointed out that the basic process
involved is defined nonrelativistically, and can be used if
relativistic effects can be safely neglected. The development
of a relativistically covariant formulation of Brownian motion could therefore
provide some insight into the structure of a relativistic quantum theory.
\par In recent years,
Brownian motion has been applied as a mechanism for the description 
of irreversible evolution of quantum mechnical states, for example,
collapse of wave functions as result of measurement.  One finds that
the wave function with
a stochastic term added to the Schr\"odinger equation (represented in
the Hilbert space of states$^2$ or in their projective
representation$^3$) evolves to a mixed state of alternative outcomes of
a measurement with the same probability
distribution as given by the calculation of {\it a priori} (Born)
probabilities in the framework of the standard quantum theory $^{2,3}$. One of the motivations for the work of Adler and
Horwitz$^3$ was based on the existence of a statistical mechanics,
developed by Adler$^{4}$ and Adler and Millard$^{5}$ for
the description of the equilibrium state of a general class of quantum
fields;  in this theory, the thermal average of commutator expressions
 take on the values implied by the familiar complex canonical quantum
theory and quantum field theory.  The Brownian motion corrections to
the standard theories may be thought of as arising from the
fluctuations around the equilibrium state. The development of a
relativistically covariant theory of Brownian motion could make an
extension of these ideas to relativistic quantum theory and quantum
field theory more accessible.
\par The program of stochastic quantization of Parisi and Wu$^6$
 assumes
the existence of relativistic wave equations, and applies a
statistical approach closely related to path integrals.  The Wiener
distribution of nonrelativistic Brownian motion, as is well known,
is associated with (imaginary time) path integral
formulations, and one would expect that covariant Brownian motion
would similarly be associated with Parisi-Wu stochastic
quantization.
\par Nelson$^1$ has pointed out that the formulation of his stochastic
mechanics in the context of general relativity is an important open
 question.  The Riemannian metric
spaces one can achieve, in principle, which arise due to nontrivial
correlations between fluctuations in spacetime directions, could,  in the
 framework of a covariant theory of Brownian motion, lead to
spacetime pseudo-Riemannian metrics in the structure of diffusion and
Schr\"odinger equations.
\par In this paper we shall study the structure a covariant theory of
Brownian motion. 
\par  We first point out some of the obvious difficulties in        
reaching a covariant theory of Brownian motion, and indicate the
directions we have chosen to solve these problems.
\par Brownian motion, thought of as a series of ``jumps'' of a
particle along its path, necessarily involves an ordered sequence.  In
the nonrelativistic theory, this ordering is naturally provided by the
Newtonian time parameter.  In a relativistic framework, the Einstein
time $t$ does not provide a suitable parameter. If we contemplate
jumps in spacetime, to accomodate a covariant formulation, a possible
spacelike interval between two jumps may appear in two orderings in
different Lorentz frames. We therefore adopt the invariant
parameter $\tau$ introduced by Stueckelberg$^7$ in his construction of
 a relativistically covariant  classical and quantum dynamics.  For
the many body theory, Piron and Horwitz$^8$ postulated that this parameter
is universal, as the Newtonian time in the nonrelativistic theory, and
drives the classical particle trajectories $x_i\,^\mu(\tau)$
(worldlines labelled $i=1,2,3,....N$)
 through equations of motion, and the evolution of the wave function
in the Hilbert space $L^2(R^{4N})$, $\psi_\tau(\{x_i\, ^\mu\})$ through the
Stueckelberg-Schr\"odinger equation (the differential form of
the action of a one-parameter unitary group with parameter $\tau$). 
\par A second fundamental difficulty in formulating a covariant theory
of Brownian motion lies in the form of the correlation function of the
random variables of spacetime. The straightforward generalization of
the usual Brownian correlation property to special relativity, i.e., 
$$ <dw_\mu(\tau)dw_\nu(\tau')>=\cases{ 0 &  $\tau \not= \tau'$ \cr 2
\alpha \eta_{\mu\nu}d\tau & $\tau=\tau'$, \cr} $$
contains the serious problem that $ <dw_0(\tau)dw_0(\tau)> <0,$ which
is impossible.  Brownian motion in spacetime, however, should be a
generalization of the nonrelativistic problem, constructed by
observing the nonrelativistic process from a moving frame according to
the transformation laws of special relativity.  Hence, as a first
step,  the process
taking place in space in the nonrelativistic theory should be replaced
by a spacetime process in which the Brownian jumps are spacelike.  The
pure time (negative) self-correlation does therefore not occur. In
order to meet this requirement, we
shall use a coordinatization in terms of generalized polar coordinates
which assure that all jumps are spacelike. In this case, one would
expect a distribution function of the form $e^{-{\mu^2\over a d\tau}}$,
 where $\mu$
is the invariant spacelike interval of the jump, and $a$ is some constant.
 As we shall see, a Brownian
motion based on purely spacelike jumps does not yield the
correct form for an invariant diffusion process.  We must therefore
consider the possibility as well that, in the framework of
relativistic dynamics, there are timelike jumps.  In a frame
in which the timelike jumps are pure time, the construction of the
Gaussian distribution from the central limit theorem can again be
applied. The distribution would be expected to be of the form
$e^{-{\sigma^2\over b d\tau}}$,
where $\sigma$ is the timelike interval of these jumps, and $b$ is
some constant.
 By suitably
weighting the occurrence of the spacelike process (which we take for
our main discussion to be ``physical'', since its nonrelativistic
limit coincides with
the usual Brownian motion) and an analytic continuation of the
timelike process, we show that
one indeed obtains a Lorentz invariant Fokker-Planck equation in
which the d'Alembert operator appears in place of the Laplace
operator of the 3D Fokker-Planck equation. One may, alternatively,
consider the timelike process as ``physical'' and analytically
continue the spacelike (``unphysical'') process to achieve a
d'Alembert operator with opposite sign.
\bigskip
\noindent
{\bf 2.  Brownian motion in 1+1 dimensions}
\smallskip
\par We consider a Brownian path in $1+1$ dimensions of the form
$$ dx^\mu (\tau) =  K^\mu(x(\tau))d\tau + dw^\mu (\tau). \eqno(2.1)$$  
\par We start by considering the second order term in the series
expansion of a function of position of the particle on the world line,
$f(x^\mu(\tau) + \Delta x^\mu)$,
involving the operator 
$${\cal O}={\Delta x}^{\mu}{\Delta x}^{\nu}{{\partial} \over {\partial}{x}^\mu}{\partial \over \partial {x}^\nu}. \eqno(2.2)$$
We have remarked that one of the difficulties in describing Brownian
motion in spacetime is the possible occurrence of a negative value for
the second moment of some component of the Lorentz four vector random
variable. If the Brownian jump is timelike, or spacelike, however, the
components of the four vector are not independent, but must satisfy
the timelike or spacelike constraint.  Such constraints can be
realized by using parametrizations for the jumps in which they
are restricted geometrically to be timelike or spacelike.
  We now separate the jumps into spacelike jumps and timelike jumps
accordingly, i.e., for the spacelike jumps,
$$ \Delta x=\mu \cosh{\alpha} \,\,\,\, \Delta t =\mu \sinh{\alpha}\eqno(2.3)$$
and for the timelike jumps,
$$ \Delta x=\sigma \sinh{\alpha} \,\,\,\, \Delta t =\sigma \cosh{\alpha} \eqno(2.3')$$
Here we assumed that the two sectors have the same distribution 
on the hyperbolic variable.
We first look for the effects of a particle experiencing spacelike
jumps only. In that case the operator ${\cal O}$ takes the following form:  
$$ {\cal O}= \mu^2[{\cosh^2}{\alpha}{\partial^2 \over \partial x^2}+2\sinh\alpha \cosh\alpha {\partial^2 \over {\partial x \partial t}}+\sinh^2\alpha{\partial^2 \over \partial t^2}] \eqno(2.4)$$
If the particle going under timelike jumps only we find the operator
${\cal O}$ takes the following form:
$$ {\cal O}= \sigma^2 [\sinh^2 \alpha{\partial^2 \over \partial
x^2}+2\sinh\alpha \cosh\alpha {\partial^2 \over {\partial x \partial
t}}+\cosh^2\alpha {\partial^2 \over \partial t^2}] \eqno(2.5)$$
\par In order to obtain the relativistically invariant d'Alembert 
diffusion operator, the expression obtained in the timelike region
must be {\it subtracted} from the expression for the spacelike
region, and furthermore, the amplitudes must be identified.  In the
physical timelike region, the coefficient $\sigma^2$ is, of course,
positive, and using the law of large numbers on the random
distribution, one obtains a Gaussian distribution analogous to that of
the spacelike case. 
\par  We see, however, that we can obtain the
d'Alembert operator only by considering the analytic continuation of
the timelike process to the spacelike domain. This procedure is
analogous to the effect, well-known in relativistic quantum scattering
theory, of
a physical process in the crossed ($t$)channel on the observed process in
the direct ($s$) channel.   Although we are dealing with an apparently
 classical
process, as Nelson has shown, the Brownian motion problem gives rise
to a Schr\"odinger equation, and therefore contains properties  of the 
quantum theory.  We thus see the
remarkable fact that one must take into account the physical effect of
the analytic continuation of processes occurring in a non-physical,
 in this case timelike, domain, on the total observed behavior of the system.
\par In the non-stochastic case, Einstein's relativity identifies
 $\Delta x /\Delta t$ with $p/E$, where $p$ and $E$ are the components
 of the energy-momentum four-vector of the particle.  If we make an
 analogous identification, assigning these variables as properties of
 the fluctuations, then 
$$ \sigma^2 = (\Delta t)^2 - (\Delta x)^2 \propto (E^2 - p^2), \eqno(2.7)$$
defining a stochastic mass squared associated with the
Brownian particle.  If the relation between $E$ and $p$ becomes
spacelike, the notion of stochastic mass can be retained undere the
transformation to an
 imaginary representation
 $E\rightarrow iE'$ and $p\rightarrow ip'$, for $E', p'$
 real\footnote{\S}{ This transformation is similar to the analytic continuation
 $p\rightarrow ip'$ in nonrelativistic tunneling, for which the
 particle appears as an instanton.}, so
 the
 relation $p'/E' >1$ remains, but $ p'^2 -E'^2 >0$. We assume
sufficient symmetry in the spacelike and timelike distributions so
that the absolute values of $<\sigma^2>$ and $<\mu^2>$ are equal.  The 
preservation of the mean
magnitude of the interval reflects the conservation of a mass-like
property which remains, as an intrinsic property of the particle, for
both spacelike and timelike jumps.
\par With these assumptions,  the cross-term in
hyperbolic functions cancels in the sum, which now takes the form
$$\langle{\cal O}\rangle = <\mu^2>\bigl[{\partial^2\over \partial x^2}
 - {\partial^2
\over \partial t^2}\bigr] \eqno(2.8)$$
Taking into account the drift term in
$(2.1)$, one then finds the relativistic Fokker-Planck equation
$$ {\partial D(x,\tau)\over \partial \tau} = \bigl\{-{\partial \over \partial
x^\mu}K^\mu + \langle \mu^2 \rangle {\partial^2 \over \partial x^\mu
\partial x_\mu}\bigr\} D(x,\tau), \eqno(2.9)$$
where $\partial/\partial x^\mu$ operates on both $K^\mu$ and $D$.
\par We see that the procedure we have followed permits us to construct
the Lorentz invariant d'Alembertian operator, as required for
obtaining a relativistically covariant diffusion equation.
Furthermore, since the expectation of $\sinh^2 \alpha, \cosh^2\alpha$
could be infinite (e.g., for a uniform distribution on $\alpha$), the
result we obtain in this way constitutes an effective regularization.
 
\bigskip
\noindent
{\bf 3. Brownian motion in  $3+1$ dimensions}
\smallskip
\par In the $3+1$ case, we again separate the jumps into timelike and
 spacelike types. The spacelike jumps may be parametrized, in a given
 frame,  by
$$ \eqalign{\Delta t =& \mu \sinh{\alpha} \cr \Delta x =& \mu \cosh{\alpha}\cos{\phi}\sin{\vartheta} \cr \Delta y =& \mu \cosh{\alpha}\sin{\phi}\sin{\vartheta} \cr \Delta z =& \mu \cosh{\alpha}\cos{\vartheta} \cr} \eqno(3.1)$$
\par We assume  the four variables $\mu, \alpha, \vartheta, \phi$ are 
independent random variables.  In addition we demand in this frame that
 $\vartheta$ and $\phi$ are uniformly distributed in their ranges
 $(0,\pi)$ and $(0, 2\pi)$, respectively. In this case, we may average
 over the
 trigonometric angles, i.e., $\vartheta$ and $\phi$ and find that:
$$\eqalign { <{\Delta x}^2>_{\phi , \vartheta}&= <{\Delta y}^2>_{\phi,\vartheta}= <{\Delta z}^2>_{\phi , \vartheta}={\mu^2 \over 3 }{\cosh}^2{\alpha}\cr  <{\Delta t}^2>_{\phi , \vartheta}=&\mu^2 {\sinh}^2 {\alpha}\cr} \eqno(3.2)$$
We may obtain  the averages over the trigonometric angles of the
timelike jumps by replacing everywhere in Eq.$(3.2)$
$$\eqalign { \cosh^2{\alpha} &\leftrightarrow \sinh^2{\alpha} \cr
\,\,\,\,
 \mu^2 &\rightarrow \sigma^2 \cr}$$ 
to obtain
$$\eqalign { <{\Delta x}^2>_{\phi , \vartheta}&= <{\Delta y}^2>_{\phi,\vartheta}= <{\Delta z}^2>_{\phi , \vartheta}={\sigma^2 \over 3 }{\sinh}^2{\alpha}\cr <{\Delta t}^2>_{\phi , \vartheta}=& \sigma^2{\cosh}^2 {\alpha},\cr} \eqno(3.3)$$
where $\sigma$ is a real random variable, the invariant timelike interval.
Assuming, as in the $1+1$ case, that the likelihood of the jumps being
in either the spacelike or (virtual) timelike phases are equal, and making an
analytic continuation for which $\sigma^2 \rightarrow -\lambda^2$,
  the total average of the operator ${\cal O}$, including the contributions of 
the remaining degrees of freedom $\mu,\lambda$ and $\alpha$  is
$$ \eqalign{<\cal O>=&\bigl(<\mu^2> <{\sinh}^2{\alpha}>-<\lambda^2><{\cosh}^2{\alpha}>\bigr){\partial^2\over \partial t^2}+\cr &{1\over 3} \bigl(<\mu^2>< \cosh^2{\alpha}>-<\lambda^2 ><\sinh^2{\alpha}>\bigr){\bigtriangleup}\cr} \eqno(3.4)$$
If we now insist that the operator $<{\cal O}>$ be invariant under 
Lorentz transformations (i.e. the d'Alembertian) we impose the condition 
$$\eqalign{ <\mu^2><{\sinh}^2{\alpha}>-<\lambda^2>&<{\cosh}^2{\alpha}>=\cr &-{1\over 3} \bigl(<\mu^2> <\cosh^2{\alpha}>-<\lambda^2><\sinh^2{\alpha}>\bigr)\cr} \eqno(3.5)$$
Using the fact that $<\cosh^2{\alpha}>-<\sinh^2{\alpha}>=1$, and
defining $ \gamma \equiv <\sinh^2{\alpha}>$, we find that
$$<\lambda^2>={1+4\gamma \over 3+4\gamma}<\mu^2> \eqno(3.6)$$
The Fokker-Planck equation then takes on the same form as in the $1+1$
case, i.e., the form $(2.9)$.
We remark that for the $1+1$ case, one finds in the corresponding expression that the
$3$ in the denominator is replaced by unity, and the coefficients $4$ are replaced by
$2$; in this case the requirement reduces to $<\mu^2> =<\lambda^2>$
 and there is no $\gamma$ dependence.

\par We see that in the limit of a uniform distribution in $\alpha$,
for which $\gamma \rightarrow \infty$, \hfil
\break $<\lambda^2> \rightarrow<\mu^2>$.  In this case, the relativistic generalization of a
nonrelativistic Gaussian distribution of the form $e^{-{{\bf r}^2\over dt}}$
becomes of the form
$e^{-{\mu^2\over d\tau}}$, which is Lorentz invariant.  As in the $1+1$
case, the result $(3.6)$ corresponds to a regularization.

\bigskip
\noindent
{\bf 4. Conclusions and Discussion}
\smallskip
\par We have constructed a relativistic generalization of Brownian
motion, using an invariant world-time to order the Brownian
 fluctuations, and separated consideration of spacelike and timelike
 jumps to avoid the problems of negative second moments which might
 otherwise follow from the Minkowski signature. Associating the
 Brownian fluctuations with an underlying dynamical process,  one may
 think of $\gamma$ in the $3+1$ case as an order
 parameter, where the distribution function (over $\alpha$),
 associated with the velocities,  is
 determined by the temperature of the underlying dynamical system (as
 we have remarked, the result for the $1+1$ case is independent of the
 distribution on the hyperbolic variable).
\par  At equilbrium, where $\partial D/\partial \tau =0,$ the
 resulting diffusion equation turns into a classical wave equation
 which, in the absence of a drift term $K^\mu$, is the wave equation
 for a massless field.  An exponentially decreasing distribution in
 $\tau$ of the form $\exp{-\kappa \tau}$ would correspond to a
 Klein-Gordon equation for a particle in a tachyonic state (mass
 squared $-\kappa$).  We have considered the spacelike jumps as
 ``physical'' since they result in the usual Brownian motion in the
 nonrelativistic limit.  If the timelike jumps were considered as
 ``physical'', one would analytically continue the ``unphysical''
 spacelike process.  The resulting diffusion equation would have the
 opposite sign for the d'Alembert operator, and an exponentially
 decreasing distribution would then result in a Klein-Gordon equation
 in a timelike particle state.

 \par  Nelson$^1$ has shown that non-relativistic Brownian motion can
 be associated with a Schr\"odinger equation. Equipped with the
 procedures we presented here, constructing relativistic Brownian
 motion, Nelson's methods can be generalized. One then can construct
 relativistic equations of Schr\"odinger (Schr\"odinger-Stueckelberg)
 type. The eigenvalue equations for these
 relativistic forms are also Klein-Gordon type equations. Moreover one can also
 generalize the case where the fluctuations are not correlated in
 different directions into the case where correlations exist, as
 discussed by Nelson$^1$ for three dimensional Riemannian spaces. In this case the resulting equation
 is a quantum equation in a curved Riemannian
 spacetime; as pointed out in ref.$10$, the eikonal approximation to the
 solutions of such an equation contains the geodesic motion of
 classical general relativity. The medium supporting the Brownian
 motion may be identified with an ``ether'' $^{11}$ (Nelson$^1$ has
 remarked that the self-interaction of charged particles
might provide a mechanism for the Brownian motion) for which the problem of
 local Lorentz symmetry is solved. This generalization of Nelson's method
 will be discussed elsewhere.
\bigskip
\noindent
{\bf Acknowledgements}
\smallskip
\par We would like to thank Linda Reichl, W.C. Schieve and Sungyan Kim
 at the Ilya
 Prigogine Center for Statistical Mechanics and Complex Systems at the
University of Texas at Austin for helpful and stimulating discussions.  

\smallskip
\noindent
{\it References}
\frenchspacing
\smallskip
\item{1.} Edward Nelson, {\it Dynamical Theories of Brownian Motion},
 Princeton University Press, Princeton (1967); Edward Nelson, {\it Quantum
Fluctuations}, Princeton University Press Princeton (1985). See also, Ph. Blanchard, Ph. Combe and W. Zheng,
{\it Mathematical and Physical Aspects of Stochastic Mechanics},
Springer-Verlag, Heidelberg (1987), for further helpful discussion.
\item{2.} For example, G.C. Ghirardi, P. Pearle and A. Rimini, Phys. Rev. A{\bf
42}, 78 (1990).
\item{3.} L.P. Hughston, Proc. Roy. Soc. London A{\bf 452}, 953
(1996), and references cited there; S.L. Adler and L.P. Horwitz, Jour. Math. Phys. {\bf 41}, 2485
(2000). 
\item{4.} S.L. Adler, Nuc. Phys. B{\bf 415}, 195 (1994); S.L. Adler,
 {\it Quaternionic Quantum Mechanics and Quantum Fields}, Oxford
Univ. Press, N.Y. (1995); S.L. Adler, hep-th/0206120.
\item{5.} S.L. Adler and A.C. Millard, Nuc. Phys. B{\bf 473}, 199 (1996).
\item{6.} G. Parisi, Y. Wu, Sci.Sin. {\bf 24} 483 ; Mikio Namiki {\it Stochastic Quantization} , (Springer-Verlag, Heidelberg, 1992).
\item{7.}  E.C.G. Stueckelberg, Helv. Phys. Acta {\bf 14}, 322 (1941); 
{\bf 14}, 588 (1941).
\item{8.} L.P. Horwitz and C. Piron,
 Helv. Phys. Acta {\bf 46}, 316 (1973).
\item{9.} See, for example,  S.S. Schweber, {\it An Introduction to
Relativistic Quantum Field Theory}, Harper and Row, N.Y. (1961).
\item{10.} L.P.Horwitz , O.Oron, hep-ph/0205018
\item{11.} S. Liberati, S. Sonego and M. Visser, Ann. Phys. 167 {\bf
298} (2002).

\vfill
\end
\bye